\documentstyle[prl,aps,epsfig,multicol]{revtex}
\begin{document}
\title{Strong coupling theory for driven tunneling and 
vibrational relaxation}
\author{M. Thorwart$^1$, M. Grifoni$^{2,3}$,  
       and 
       P. H\"anggi$^1$}
\address{
              $^1$Institut f\"ur Physik,
              Universit\"at  Augsburg, Universit\"atsstra{\ss}e 1, D-86135 Augsburg,
              Germany\\
 $^2$Institut f\"ur Theoretische Festk\"orperphysik,  Universit\"at
 Karlsruhe, D-76128
 Karlsruhe, Germany\\ 
 $^3$Dipartimento di Fisica, INFM, Via Dodecaneso 33, I-16146, Genova, Italy\\  
 }
\date{Date: \today}
\maketitle
\begin{abstract}
We investigate on a unified basis tunneling and vibrational relaxation in driven
dissipative multi-stable systems described by their $N$ lowest lying unperturbed
levels. By use of the {\it discrete variable representation} we derive a set of
coupled non-Markovian master equations. We present {\it analytical  
treatments} that describe the dynamics 
in the regime of {\em strong} system-bath coupling. 
Our findings are corroborated by 
``ab-initio'' real-time path integral calculations. 
\end{abstract}
\vspace{-3mm}
\pacs{PACS: 05.30.-d, 05.40.-a, 33.80.Be, 82.20.Pm}
PACS: 05.30.-d, 05.40.-a, 33.80.Be, 82.20.Pm 
%---
\raggedcolumns
%---
\vspace{-2mm}
\begin{multicols}{2}
%---
\narrowtext
%---
\noindent
Dissipative tunneling  in bistable systems  
finds wide\-spread applications in many 
physical and chemical contexts. Usually, the dynamics is described 
by restricting a double-well potential to 
its lowest tunneling doublet of energy eigenstates \cite{PR}. This  
two-level system (2LS) describes well the 
dynamics at low temperature, but becomes increasingly invalid at 
higher temperatures, when higher lying doublets are populated.  
Then both tunneling between the metastable quantum wells and 
vibrational relaxation within the wells mix the dynamics. 
In the presence of an external field, e.g., 
a laser, the different doublets are additionally coupled. 
Such dissipative {\em multilevel} systems occur in the context of 
 tunneling of magnetization in organic high-spin molecules, 
 i.e., 
Mn$_{12}$ \cite{Wernsdorfer99} and Fe$_8$ \cite{Sangregorio}. 
Moreover, they mimic tunneling 
of defects in metals \cite{Cukierdef} and hydrogen pair transfer in 
benzoic acid \cite{Stoeckli}. 
Also, resonant tunneling of the magnetic flux in SQUIDS exhibits  
a multilevel structure \cite{Han}. 
An example, 
where a picosecond laser driving accelerates intramolecular 
isomerization in malonaldehyde, is given in 
\cite{Kuehn}. 
There, a pumping mechanism  for fast hydrogen transfer (``hydrogen subway'') 
is  proposed (see also \cite{MorilloDDS}). 
 This complex driven dissipative multilevel system has been  
 investigated {\em numerically} in \cite{Kuehn,Dittrich}
 within a {\em weak coupling} approach,  and  
 in \cite{Thorw} within  the 
 real-time  path integral tensor multiplication scheme
 \cite{QUAPI}. 
 In absence of driving, the dissipative  bistable system has 
 been  researched by a numerical reactive flux analytic continuation 
 scheme in \cite{Berne}. 
 The only {\em analytical} approach previous to our work 
 has been attempted in \cite{MorilloDDS}. In these works, however,
  the dynamics {\em restricted} to the {\em two} lowest tunneling doublets 
  was investigated  in the {\em weak coupling} 
 regime. Because of its complexity, even the {\em nondriven} multistate dynamics 
 has barely been  studied \cite{Dekker,Parris}. 
 Common to all prior works \cite{MorilloDDS,Dekker,Parris} is that 
 the position operator is incorrectly assumed to be 
 diagonal in the localized (spin) representation. This, however, holds true
 only when  the lowest  
 tunneling-doublet rules the dynamics, cf.\  (\ref{ext}) below. 
 Moreover, the  harmonic approximation 
 for the two wells  proposed in \cite{Dekker}  
 can be justified only for a large barrier height 
 where a semiclassical description is already applicable \cite{Weiss93}.  
 Furthermore, the research in \cite{Parris} is not completely
 based on a full microscopic Hamiltonian. \\
 The prime objective of this work is the development of a {\em first consistent 
 treatment} wherein tunneling (TU), driving,  
 and vibrational relaxation (VR) are treated {\em analytically} on a 
 common footing, {\em beyond a Redfield-approach}, and for an arbitrayry 
 number $N$ of levels. A unified, 
 consistent treatment becomes possible only if one uses  
  the so termed  {\it discrete variable 
 representation (DVR)} \cite{Harris}. It is in this DVR basis 
 that the position operator  is diagonal.  
 A real-time path integral approach is used to 
 derive a set of coupled generalized master 
 equations 
 for  the combined VR and TU dynamics.
 The   predictions of these equations well agree with  
 precise numerical quantum simulations.  
 Analytical results for the  rate  matrix 
 of a $N$-level system 
 are obtained.  
\\
[0.1cm]
{\it The model.--}
 To start, we consider the Hamiltonian 
 $H(t)=H_0+
 H_{\rm ext}(t)+H_{\rm B}$  which accounts 
 for quantum 
 dissipation and external, generally time-dependent control fields.
 The  term  $H_0=\frac{p^2}{2M}+V(q)$ denotes the Hamiltonian 
 of the isolated 
 system. We consider 
 a particle of mass $M$ moving  in the bistable quartic potential 
 $V(q)=(M^2\omega_0^4/64 E_B) q^4-(M\omega_0^2/4)q^2$,  
 where $E_B$ is the barrier height, and $\omega_0$ the frequency of classical
 oscillations around the potential minima at $\pm q_0/2$.  
 For the isolated  system the energy spectrum follows from the 
 Schr\"odinger equation  $H_0|n\rangle =E_n|n\rangle$,
  $n=1,2,...$. 
 For energies well below the barrier, 
 the spectrum consists of a ladder of  doublets. 
 To illustrate the method, we consider  
 the case in which only two doublets  
 $\hbar\Delta_1=E_2 -E_1$, and $\hbar\Delta_2=E_4-E_3$ contribute 
 significantly to the  dynamics. They are separated by the energy 
 gap $\hbar \overline{\omega}_0 = \frac{1}{2} (E_4+E_3) - \frac{1}{2} 
(E_2+E_1) \gg \hbar \Delta_i$.
 We consider then the reduction $H_0\to H_{\rm 4LS}$, with 
 $H_{\rm 4LS}$ being the  
 Hamiltonian for the isolated  four-level  system (4LS). 
 The external field is characterized by
 $H_{\rm ext}(t)=-s(t)  q$, 
 with $s(t)$  being a time-dependent field. 
 In the basis of the 
 vectors  $|R_1\rangle=\frac{1}{\sqrt 2}(|1\rangle + |2\rangle)$,
 $|L_1\rangle=\frac{1}{\sqrt 2}(|1\rangle - |2\rangle)$,
 and $|R_2\rangle=\frac{1}{\sqrt 2}(|3\rangle + |4\rangle)$,
 $|L_2\rangle=\frac{1}{\sqrt 2}(|3\rangle - |4\rangle)$,
 with $|L_i\rangle$ ($|R_i\rangle$) localized in the left 
 (right) well, 
 the discrete position operator of the system  reads 
\vspace{-1mm}
\begin{eqnarray}
  q^{loc} & = & \sum_{i,j=1,2}a_{ij}(|R_i\rangle\langle R_j| -
 |L_i\rangle\langle L_j|)  \nonumber \\ & & + {} b(|L_1\rangle\langle R_2| 
+  |R_2\rangle\langle L_1| \nonumber \\ & & - {} |R_1\rangle\langle L_2|-|L_2\rangle\langle R_1|),
 \label{ext}\vspace{-1mm}
\end{eqnarray}
with  $a_{11}=\langle 1| q| 2 \rangle$,  
 $a_{22}=\langle 3| q| 4 \rangle$, $a_{12}=
 a_{21}=(\langle 1| q| 4 \rangle 
 + \langle 2| q| 3 \rangle)/2$, and 
 $b=(\langle 1| q| 4 \rangle - \langle 2| q| 3 \rangle)/2 \ll a_{ij}$. 
 Note that, in clear contrast to the 2LS  case, it is {\em nondiagonal}. 
 For the following analytical treatment, we set $b=0$. 
 In all results shown in the figures, however, we use $b\ne 0$. 
  Finally,  we model
 quantum  dissipation by an ensemble of harmonic oscillators that are  bilinearly
 coupled to the  system \cite{Weiss93}, i.e.,
$H_{\rm B}=\frac{1}{2}\sum_{i}\Big [\frac{p_i^2}{m_i}+m_i\omega_i^2
\Big (x_i-\frac{c_i}{m_i\omega_i^2} q\Big ) ^2 \Big ],$  
 with $J(\omega)=(\pi/2)\sum_i c_i^2/(m_i\omega_i)\delta(\omega-\omega_i)$  
 being the spectral density.  We assume 
 Ohmic dissipation with an exponential cut-off $\omega_c\gg\omega_0$
 and a viscous friction strength $\gamma$, 
 such that  $J(\omega)=M\gamma\omega e^{-\omega/\omega_c}$. 
 We  wish to evaluate the  probability
 $P_{\rm L}(t):= \sum_i
 \langle L_i| \rho(t)|L_i\rangle$  to find at time $t>t_0$ the
 system in the  left well.
 Here $\rho(t)$ denotes the 
 reduced density matrix (RDM) of the system.
 We  belabor the 
 case in which the particle initially is prepared in the lower
 left  state: $\rho(t_0)=|L_1\rangle\langle L_1|$, 
 with the bath in  thermal equilibrium at
 temperature $T=(k_B\beta)^{-1}$. \\
[0.1cm]
{\em {DVR} description of vibrational relaxation and  tunneling.--}\\
 For a harmonic bath,  path integral techniques allow one to trace out 
 analytically the bath degrees of freedom 
 in the eigenbasis of the position operator $q$ (DVR basis).  
 Let us start with the introduction of the DVR vectors:
\vspace{-1mm}
\begin{eqnarray}
| \alpha_1 \rangle &=&v(|L_1\rangle - u | L_2 \rangle)
\;,\qquad
| \beta_1 \rangle =v(|R_1\rangle - u | R_2 \rangle)
\;,\nonumber \\
| \alpha_2 \rangle &=&v(u|L_1\rangle +| L_2 \rangle)
 \;,\qquad 
| \beta_2 \rangle =v(u|R_1\rangle +| R_2 \rangle)
\;,\vspace{-1mm}
\end{eqnarray}
 with $| \alpha_i \rangle$  ($| \beta_i \rangle$)  being
 localized in the
 left (right) well, respectively. 
 Here, $v=1/\sqrt{1+u^2}$ and $u=(a_{11}+\lambda_{\alpha_1})/a_{12}=-(a_{22}+
 \lambda_{\alpha_2})/a_{12}$, 
 and $\lambda_{\alpha_i}=-\lambda_{\beta_i}$ denote the position eigenvalues: 
 $\lambda_{\alpha_{1,2}}=[-(a_{11}+a_{22})\mp \sqrt{
 (a_{11}-a_{22})^2+4a_{12}^2}]/2$. 
 Upon the introduction of the 
 DVR {\em tunneling}  matrix elements  
\vspace{-1mm}
\begin{eqnarray}
\Delta_{\alpha_1\beta_1} &\equiv &v^2(\Delta_1+u^2\Delta_2)\;,
\; \Delta_{\alpha_2\beta_2}\equiv v^2(u^2\Delta_1+\Delta_2)\;,
\; \nonumber\\
\Delta_{\alpha_1\beta_2}& =& \Delta_{\alpha_2\beta_1} \equiv v^2u(\Delta_1-\Delta_2)\;,
\label{delta}\vspace{-1mm}
\end{eqnarray}
 being a {\em linear combination} of the tunneling splittings $\Delta_1$ and
 $\Delta_2$, 
the  4LS Hamiltonian reads in this DVR basis  
\vspace{-1mm}
\begin{eqnarray}
\lefteqn{H_{\rm 4LS}^{DVR} = -\sum_{i,j=1,2}
\frac{1}{2}\hbar\Delta_{\alpha_i\beta_j}
(|\alpha_i\rangle\langle \beta_j|+|\beta_j\rangle\langle \alpha_i|)}
  \nonumber\\& +&  
\sum_{i=1,2}(F_{\alpha_i}|\alpha_i\rangle\langle \alpha_i|+
F_{\beta_i}|\beta_i\rangle\langle \beta_i|)  
 - v^2u\hbar\overline{\omega}_0 R \;,
 \label{HamDDSdvr}
 \end{eqnarray} 
 with $F_{\alpha_1}=F_{\beta_1}=u^2v^2\hbar\overline{\omega}_0$,  
 $F_{\alpha_2}=F_{\beta_2}=v^2\hbar\overline{\omega}_0$.  
 The matrix 
 $R=|\alpha_1\rangle\langle \alpha_2|+|\alpha_2\rangle\langle 
 \alpha_1|+|\beta_1\rangle\langle \beta_2|+|\beta_2\rangle\langle \beta_1|$
 accounts for intrawell transitions. 
 It is suggestive  to  introduce  the DVR {\em intrawell} transition elements 
\vspace{-1mm}
\begin{equation}
 \Delta_{\alpha_1\alpha_2}=\Delta_{\alpha_2\alpha_1}=
 \Delta_{\beta_1\beta_2}=\Delta_{\beta_2\beta_1} \equiv 2v^2u\overline{\omega}_0\;.\vspace{-1mm}
\end{equation}
Because  
$\Delta_{\alpha_1\beta_2} \approx \Delta_{\alpha_1\beta_1}\le 
\Delta_{\alpha_2\beta_2}< 2v^2u \overline{\omega}_0$,
 different time scales determine this VR  
 assisted tunneling dynamics. For the left well population, one finds
 $P_{\rm L}(t)=\sum_i\langle\alpha_i 
 |\rho(t)|\alpha_i\rangle$,
  with  the initial RDM  
 $\rho(t_0)=|L_1\rangle\langle L_1|=v(|\alpha_1\rangle\langle \alpha_1|
 +u^2|\alpha_2\rangle\langle \alpha_2| +u
 |\alpha_1\rangle\langle \alpha_2|+u |\alpha_2\rangle\langle \alpha_1| )$
 being {\em nondiagonal}. By use of the notation $\mu = 
 {\alpha_1,\alpha_2,\beta_1,\beta_2}$ for the index of the 
 DVR states, 
 the formally exact path integral expression for the diagonal elements of the 
 RDM  in the DVR basis reads ($\rho_{\mu \nu}:=\langle \mu| \rho(t)|\nu\rangle$)
\vspace{-1mm}
\[ \rho_{\nu \nu} (t) = \sum_{\mu_0\nu_0}
 \int {\cal D}q\int{\cal D}q'{\cal A}[q]
 {\cal A}^*[q']
{\cal F}[q,q']\rho_{\mu_0\nu_0}(t_0) \;,
\vspace{-1mm}
\]
 with the paths 
 subject to the constraint $q(t)=q'(t)=\lambda_{\nu}$, 
 and  $q(t_0)=\lambda_{\mu_0}$, $q'(t_0)=\lambda_{\nu_{0}}$, with 
 $\{ \lambda_{\nu} \}$ being the position operator eigenvalues. 
 ${\cal A}[q]$  is  the path weight in the absence of 
 dissipative forces, while   the influence functional ${\cal F}[q,q']$  
  accounts  for the bath effects. 
  \\
[0.1cm]
{\em $N$-level dynamics.-- } Although derived for a $4$LS, 
these equations hold for a finite number $N$ of levels. 
 We switch to  center of mass 
 $\eta(s)=[q(s)+q'(s)]$ and relative coordinates $\xi (s)=[q(s)-q'(s)]$. 
 Then, the double path integral over
 the ``$N$-state'' paths $q(s)$ or $q'(s)$  is 
 expressed as a {\em single} path integral over the $N^2$ states
 of the RDM in the $(q,q')$ plane. The case of the $4$LS is
  depicted in Fig.\  1.
 Any  such path  can be 
 described as a sequence of time intervals spent in a diagonal state of the RDM 
 (``sojourns'') and time intervals between two successive visits of the 
 diagonal states (``clusters''). 
 The  functional {$\cal F$} couples different path segments.
 For a path with $n$ transitions at times $t_1,..,t_n$, we 
 introduce the cumulative off-diagonal charge $p_j=\sum_{i=1}^{j \le n}\xi_i$,
   associated to the path derivative $\dot \xi (s):=\sum_{i=1}^n
  \xi _i \, \delta(s-t_i)$. With 
$\xi_i=(\lambda_{\mu_i}-\lambda_{\nu_i})-(\lambda_{\mu_{i-1}}-
   \lambda_{\nu_{i-1}})$,   it yields   
  $p_n=0$ within each cluster of $n$ jumps. 
 Thus, the clusters are ``neutral'' objects,
 being only weakly interacting
 at high temperature and/or large Ohmic damping, 
 due to the short range of the intercluster interaction. 
 This suggests a  generalization of the 
 {noninteracting-cluster-approximation} (NICA) 
 scheme in \cite{egger} to our driven  situation, with 
 combined VR and  TU  transitions also  
 among non-nearest neighbors and with an initially nondiagonal RDM. We call  
 this the  VRTU-NICA. It yields   
 the set of coupled generalized master equations (GME) 
\vspace{-1mm}
\begin{equation}
\!\!\!\!\!\!\! \dot\rho_{\nu \nu}(t,t_0)= I_{\nu}(t,t_0)+
\sum_{\mu=1}^N \int_{t_0}^{t}\! dt'H_{\nu \mu}(t,t')
 \rho_{\mu \mu}(t',t_0) \;, 
\label{gme}\vspace{-1mm}
\end{equation}
 where  all intercluster 
 as well as non-nearest-neighbors sojourn-cluster interactions 
 are neglected, while keeping 
 fully the intracluster interactions. 
The VRTU-NICA is expected to yield reliable results as long as the friction 
strength does not exceed level broadening among neighboring doublets 
or the temperature is not too low.
  For the case of a  single tunneling doublet the DVR basis 
 coincides with the localized basis. Then,
 the VRTU-NICA  reduces 
 to the familiar (driven) noninteracting-blip approximation
 for a 2LS  \cite{PR}.
 If  more than  two levels are involved, 
 the functions $H_{\nu \mu}$ and $I_{\nu}$ are  expressed as power series 
 in the DVR-Hamilton matrix elements  $\Delta_{\mu \nu}$.  
  \\
[0.1cm]
{\em Sequential dynamics.--} 
 To lowest order one finds  
\vspace{-1mm}
 \begin{eqnarray} H^{}_{\nu\mu}(t,t')&=&\frac{\Delta^2_{\nu\mu}}{2}
 e^{-Q_{\nu\mu}'(t-t')} \cos [ \varphi_{\nu\mu}(t,t')
 -Q_{\nu\mu}''(t-t')]\;,\nonumber\\
 I_{\nu}(t,t_0)& =& (\delta_{\nu\alpha_2}-\delta_{\nu\alpha_1} 
 ) {\rm Re} \rho_{\alpha_1\alpha_2}(t_0) 
 \Delta_{\alpha_1\alpha_2} 
 e^{- Q_{\alpha_1\alpha_2}'(t-t_0)} \nonumber \\
 &  \times &
 \sin[\varphi_{\alpha_1\alpha_2}(t,t_0)-
 Q_{\alpha_1\alpha_2}''(t-t_0)]\;,
 \; \nu \ne \mu .
 \label{kernel2}
\vspace{-1mm}\end{eqnarray}
 The conservation of probability yields for 
 the diagonal elements $H_{\mu\mu}= - \sum_{\nu \ne \mu} H_{\nu \mu}$.
 The driving influence is captured by 
$\varphi_{\nu \mu}(t,t')=\int_{t'}^{t} dt''[\varepsilon_\mu (t'') -
 \varepsilon_\nu(t'')]$, where 
$\varepsilon_\nu (t)=F_\nu - \lambda_{\nu} s(t)$ with $F_\nu$ given below 
(\ref{HamDDSdvr}).  
 Finally, the bath influence  is encapsulated in  the 
 correlation functions $Q_{\mu\nu}=Q_{\mu\nu}'+iQ_{\mu\nu}''$ with
 $Q_{\mu\nu}=(\lambda_{\mu}-\lambda_{\nu})^2 Q(t)$, and 
  $Q_{}(t)=\frac{1}{\hbar\pi}\int_{0}^{\infty}d\omega
 [J(\omega)/\omega^2]
 (\{\cosh (\beta\hbar\omega /2)-\cosh[\omega(\frac{\hbar\beta}{2}-it)]\}/
 [\sinh (\beta\hbar\omega/2)])$.   
 Hence, to each transition an {\em effective}  friction strength
 \begin{equation} 
\alpha_{j}\equiv (\xi_j/q_0)^2\alpha, \quad  \mbox{with} \quad
  \alpha=M\gamma q_0^2/(2\pi\hbar)\, , 
\end{equation} 
 is associated.
 The prediction for the population $P_{\rm L}(t)$ of the
 left well is depicted 
  in the inset of Fig.\ 2, together 
 with the results of the
 quasiadiabatic propagator 
 path integral method (QUAPI) \cite{QUAPI} adapted to the 4LS 
  case. For the chosen parameters,    
   higher order coherent paths yield only minor corrections. 
 The inset also shows that the dynamics 
 described by (\ref{gme}) is  well approximated by a
 Markovian master equation, being {\em independent} 
 of the initial off-diagonal preparation, i.e., 
 $ \dot\rho_{\nu \nu}(t)=\sum_{\mu} \Gamma_{\nu \mu}(t)
 \rho_{\mu \mu}(t)\;,$
 where  $\Gamma_{\nu \mu}(t)=\int_{0}^{\infty}dt'H_{\nu \mu}(t,t-t')$.
  The explicit time dependence of $\Gamma_{\nu\mu}$ reflects the 
 time dependent external forcing.
 It is appearent that the dynamics
 is in this regime governed by a {\em single exponential decay}.
 To extract this decay rate we observe that
 for  high-frequency fields (such as an interdoublet  
 resonant field), averaging over a driving period is appropriate.
 This yields the time-independent rate-matrix 
\vspace{-2mm}
\begin{eqnarray}
\Gamma_{\nu\mu}^{\rm av}& = &\frac{\Delta_{\nu\mu}^2}{2}
\int_{0}^{\infty} d\tau \exp[-Q_{\nu\mu}'(\tau)] 
J_0\Big(\zeta_{\mu\nu}\frac{2s}{\Omega} \sin\Big(\frac{\Omega\tau}{2}\Big)\Big)
\nonumber\\ &\times&
\cos [ (F_{\mu} - F_{\nu}) \tau
 -Q_{\nu\mu}''(\tau)]\; ,\;\;\nu \ne \mu\;, \label{avrate}\vspace{-1mm}
\end{eqnarray}
 where $\zeta_{\mu\nu}=\lambda_{\mu}-\lambda_{\nu}$, with $J_0(x)$ being the zeroth Bessel function. 
 The main part of Fig.\  2 
 shows the 4LS averaged decay rate $\Gamma^{\rm av}$,
 being the {\em smallest nonzero eigenvalue} of the Markovian
 rate matrix $\Gamma_{\nu\mu}^{\rm av}$, 
vs the amplitude $s$ of a resonant ($\Omega= \overline{\omega}_0$) 
 driving field $s(t)=s \sin(\Omega t)$. 
 Note the perfect agreement between the GME predictions and those
 of the QUAPI, together with the  
  characteristic {\it nonmonotonic} behavior. 
 The important issue of the contribution of the higher lying states
 is investigated in Fig.\ 3, where the undriven  
 ($s=0$) decay rate is depicted vs  the
 number $N$ of DVR states used to truncate the full double-well
 problem. It is clearly seen that, for moderate friction,
  obeying $\Delta_i < \gamma < \overline{\omega}_0$, 
 the truncation to  the lowest two doublets is adequate  even 
 at moderate temperature $k_BT=0.1 \, \hbar\omega_0$. 
 This condition implies that neighboring doublets do not overlap due to frictional broadening. 
 Clearly, the convergence is expected to improve as the temperature is lowered.  
 We find (not shown) that a truncation to a 4LS is adequate for 
 {\em low} temperatures and {\em slow} driving fields ($\Omega < 0.1 \, \omega_0$).
 For  high-frequency fields  also higher lying states are involved in the
 dynamics as depicted in the
 inset of Fig. 3. Hence, a reduction of the driven double-well problem to
 a 4LS is problematic in the presence
 of a strong resonant field and moderate temperatures.  
\\
[0.1cm]
{\it Beyond sequential dynamics.--}
Upon increasing the temperature ($k_BT\stackrel{>}{\sim}\hbar\omega_0$), 
 a truncation to a few levels only starts to be inadequate.  
Because the effective friction strengths $\alpha_{j}$
 scale quadratically with  $\xi_j$,
 upon increasing the number $N$ of DVR states involved,  the $N$LS effectively 
 {\em flows to weak coupling}. For small effective $\alpha_{j}$, 
 however, the noise action does not
  suppress long intervals in the off-diagonal states, and the
 higher order 
 paths start to contribute.
 For $\gamma < 0.1 \, \omega_0$, $k_BT\stackrel{>}{\sim}\hbar \omega_0$  
 and $\omega_c \gg  k_BT/\hbar, \omega_0$,
  we can approximate 
 $Q(t)\approx 2\alpha[\pi |t|/\hbar\beta +\ln(\hbar\beta\omega_c/2\pi)]
 +i\pi\alpha 
 \;{\rm sgn} \, t$.  
 Now the intercluster correlations cancel out  exactly. 
  The corresponding averaged Markovian rates read
 \begin{eqnarray}
 \lefteqn{\!\!\!\!\!\!\!\!
\Gamma_{\nu\mu}^{\rm av}=\sum_{n=2}^\infty \sum_{\{\mu_j\nu_j\}}
 \prod_{j=1}^n (-1)^{\delta_{j}}  \Big(\frac{i}{2}\Big)^n   
 \tilde{\Delta}_{\mu_j \nu_j,\mu_{j-1} \nu_{j-1}} 
 \Big(\frac{2\pi}{\beta\hbar\omega_c}\Big)^{\alpha_j }} \nonumber\\
& & \times \; 
 e^{-i\pi (-1)^{\delta_j} \alpha_jp_j/\xi_j}f_j\;, 
 \label{ratemarkov}\\
 f_j &=&
\int_0^\infty d\tau 
 e^{-\frac{2\pi\alpha_jp^2_j }{\hbar\beta \xi^2_j}\tau } 
 J_0\Big(p_j\frac{2s}{\Omega}
 \sin\Big( \frac{\Omega\tau}{2}\Big)\Big)
 e^{ - i(F_{\mu_j}-F_{\nu_j})\tau}\; , \nonumber 
\end{eqnarray}
where $p_j=\sum_{i=1}^{j\le n}\xi_i$, $\delta_j=0 \, (1)$ 
for a vertical (horizontal)
jump and 
$\tilde{\Delta}_{\mu_j \nu_j,\mu_{j-1} \nu_{j-1}}$ 
is the DVR-Hamilton matrix element for the transition from 
$(\mu_{j-1}, \nu_{j-1})$ to $(\mu_j, \nu_j)$, e.g.\  for a horizontal jump, 
$\tilde{\Delta}_{\mu_j \nu_j,\mu_{j-1} \nu_{j-1}} := 
\Delta_{\nu_j \nu_{j-1}}$. Since 
(\ref{ratemarkov}) describes well the 
 high temperature dynamics, this rate matrix
  with $s=0$ constitutes the starting point
 for an evaluation of the crossover to the classical regime. 
 With $s\neq 0$, however, a chaotic dynamics may occur.  \\
{\it Conclusions.--} 
 We put forward a real-time path 
 integral approach to investigate 
 the interplay between vibrational relaxation  and   tunneling in  
 driven, dissipative  multilevel quantum systems. We  succeeded
 in deriving a novel GME  within the DVR basis 
 which treats  tunneling and intrawell dynamics  
 on a common basis: Its Markovian approximation in (\ref{avrate}) and  
 (\ref{ratemarkov}) yields novel 
 analytical results that well agree with those of the full GME. 
 In contrast to   semiclassical imaginary-time rate calculations
 \cite{Weiss93}, 
 we are not  limited by the requirement of 
 thermal equilibrium at adiabatically varying external fields.
 Hence, our results provide a 
 powerful tool to investigate the crossover from a 
 quantum to a classical dynamics  beyond the restrictions of
 the semiclassical approximation. 
Our choice of parameter finely mimics the situation of 
a picosecond laser which accelerates 
 isomerization in malonaldehyde \cite{Kuehn}.
There a barrier height of $E_B \approx 1.7$ is given,
to be contrasted with $E_B=1.4$ here. 
Also ${\rm Mn}_{12}$ and $\rm{Fe}_{8}$ nanomagnets,
 which have a spin ground state of $S=10$ ($N=21$ levels), provide
 an interesting example where to apply our theory
 \cite{Wernsdorfer99}.
\vspace{-1mm}
\centerline{* * *}
We acknowledge support  by the Deutsche
Forschungsgemeinschaft (HA1517/14-3, P.H.),  
 the Graduiertenkolleg GRK 283 (M.T.), 
and the European Community (M.G.). 
We thank 
I.\ Goychuk, 
G.\ Ingold, P.\ Reimann,  and G.\ Sch\"on  for fruitful remarks. 
\vspace{-7mm}
\begin{figure}[th]
\begin{center}
\epsfig{figure=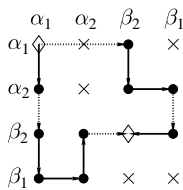,width=25mm,height=25mm,angle=0} 
\end{center}
\vspace{-1mm}
\caption{The 
sixteen states of the reduced density matrix 
 of a four-level system (4LS) in the discrete 
 variable representation.
 Shown are 
 two typical  paths 
 starting and ending in the diagonal states $\Diamond$
 Solid lines indicate vibrational relaxation 
 transitions,  dotted lines tunneling events.}
\end{figure}
\vspace{-3mm} 
\noindent 
\begin{figure}[th]
\begin{center}
\epsfig{figure=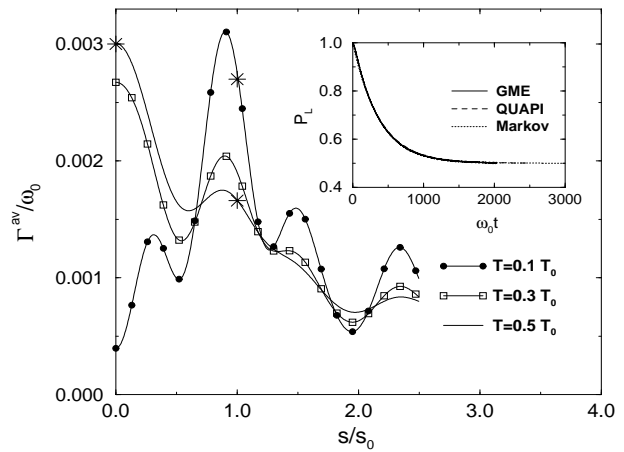,width=8cm,height=6cm,angle=0}
\end{center}
\vspace{-3mm}
\caption{Averaged transfer rate of a 4LS 
  vs.\  the scaled ($s_0=\hbar\omega_0/q_0$, $q_0=\sqrt{\hbar/M\omega_0}$) 
 driving amplitude of a resonant  
 ($\Omega=\overline{\omega}_0=0.815 \, \omega_0$) ac-field $s\sin\Omega t$  
 as obtained from 
 the smallest nonzero eigenvalue of the rate matrix  
 (\protect\ref{avrate}). 
 The three asterisks $\ast$ denote findings of an exponential fit to the 
 QUAPI  results. Note the  resonant  enhancement 
 at finite driving strength as the  temperature is lowered  
 ($T_0=\hbar\omega_0/ k_B$). Here and in the inset 
 we set $E_{\rm B}=1.4 \, \hbar\omega_0$, $\gamma=0.1 \, \omega_0$,  
 $\omega_c=10 \, \omega_0$. Inset: Evolution of the population of
 the left well as predicted by the GME (\protect\ref{gme}) 
 with (\protect\ref{kernel2}), by its Markovian approximation and by the 
 quasi-adiabatic path integral method (QUAPI). We 
 choose $T=0.1 \, T_0$, $s=1.0 \, s_0$.} 
\end{figure}
\begin{figure}[th]
\begin{center}
\epsfig{figure=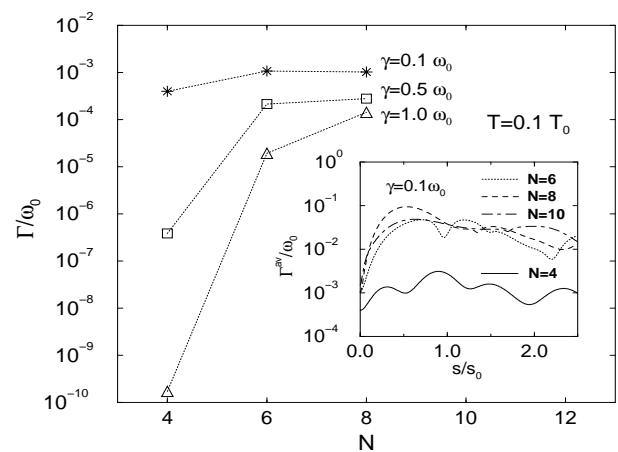,width=8cm,height=6cm,angle=0}
\end{center}
\vspace{-3mm}
\caption{Static deacy  rate  $\Gamma$ 
  vs.\ the number $N$ of levels used to truncate the full
 double-well dynamics. Even at moderate temperatures $T=0.1 \, T_0$ the 
 reduction to the two lowest doublets is appropriate for moderate damping.
 Here and in the inset we set $E_{\rm B}=1.4 \, \hbar\omega_0$,  
 $\omega_c=10 \, \omega_0$.
   Inset: Averaged transfer rate $\Gamma^{\rm av}$ vs.\  
  driving strengths $s$ of a resonant field
 ($\Omega=\overline \omega_0$) at $T=0.1 \,  T_0$.
 Convergence requires now $N>4$.} 
\end{figure}
\vspace{-3mm}
\noindent 
%
%---
\end{multicols}
\end{document}